\documentclass[12pt,preprint]{aastex}

\slugcomment{This manuscript is prepared for submission to
\textbf{MNRAS}}

\shorttitle{Star formation} \shortauthors{Gholipour and
Nejad-Asghar}

\begin{document}

\title{Rossby-wave instability in viscous discs}

\author{
Mahmoud Gholipour and Mohsen Nejad-Asghar}

\affil{Department of Atomic and Molecular Physics, University of
Mazandaran, Babolsar, Iran} \email{mhd.gholipour@stu.umz.ac.ir}

\begin{abstract}
The Rossby wave instability (RWI), which depends on the density
bumps and extremum in the vortensities in the differentially
rotating discs, plays an important role in the evolution of the
protoplanetary discs. In this article, we investigate the effect of
viscosity on the non-axisymmetric RWI in the self-graviting
accretion discs. For this purpose, we add the viscosity to the work
of Lovelace and Hohlfeld (2013). Consideration of viscosity
complicates the problem so that we use the numerical method to
investigate the stable and unstable modes. We consider three ranges
of viscosities: high viscosity in the ranges $0.1\leq \alpha\leq
0.4$, moderate viscosity in the ranges $0.01\leq \alpha < 0.1$, and
low viscosity in the ranges $\alpha < 0.01$. The results show that
the occurrence of the RWI is related to the value of viscosity so
that the effect of high viscosity is important, while the low
viscosity is negligible. These results may be applied for the study
of the RWI role in planet formation and angular momentum transport
for different kinds of the protoplanetary discs with different
viscosities.
\end{abstract}

\keywords{accretion, accretion discs —- instabilities —-
hydrodynamics —- waves —- spiral galaxies}

\section{Introduction}
The theory of Rossby wave instability (RWI) was developed by
Lovelace et al. (1999) and Li et al. (2000) for thin accretion discs
with negligible viscosity and self-gravity (Lovelace \& Hohlfeld
2013). The criterion to obtain the RWI depends on the density bump
and an extremum in the vortensity. A density bump could arise at the
radial boundary of the dead zone (e.g., Tagger \& Varniere 2006;
Lyra et al. 2009; Ataiee et al. 2013) and vortensity is defined by
$\zeta=\kappa^2/(2\Omega \Sigma)$, where $\Sigma$ is the disc
surface density, $\Omega$
 is the disc rotation rate and $\kappa^2=4\Omega^2+2r\Omega d\Omega/dr$
 is the square of the radial epicyclic frequency
 (e.g., Narayan et al. 1987; Drazin \& Reid 2004; Varniere \& Tagger 2006; Yu \& Lai 2013). For two-dimensional (vertically
integrated) barotropic discs, the RWI relies on the existence of an
extremum in the background fluid vortensity. Since Rossby waves
propagate along the gradient of vortensity, the instability can be
understood as arising from the interaction between two Rossby waves
propagating on each side of the vortensity extremum (Yu \& Lai
2013).

Recently, the RWI has  been reviewed and developed by some
astronomers.  For example, Lovelace and Hohlfeld (2013) analyzed the
RWI in a continuum of thin disc models ranging from self-gravitating
to non-selfgravitating cases. They found that the important
quantities determining the stability/instability are: (1) the
parameters of the bump (or depression) in the inverse potential
vortensity (2) the Toomre $Q$ parameter of the disc, and (3) the
dimensionless azimuthal wavenumber of the perturbation
$\bar{k}_\phi= mQh/r_0$, where $r_0$ represents the bump place, $h$
is the half-thickness of the disc and $m = 1, 2, ...$ is the
azimuthal mode number. As another study, we can mention the work of
Yu and Lai (2013) in which they studied the effect of large-scale
magnetic fields on the non-axisymmetric RWI in the accretion discs.
They show that the instability develops around a density bump, which
is likely present in the transition region between the active and
dead zones of the protoplanetary discs. Some of the simulations on
the RWI have been performed with considering the viscosity (e.g.,
Varniere \& Tagger 2006), and some of them have been done in the MHD
resistive discs (e.g., Lyra \& Mac Low 2012).

The viscosity has a significant impact on the evolution of the thin
discs. The standard model of the viscous accretion discs was
formulated in the well known papers of Shakura~(1972) and Shakura
and Sunyaev~(1973). The viscosity affects the density bump, which is
likely present in the transition region between the active zone and
dead zone of the protoplanetary discs. The viscous torque at the
transition has a component proportional to the negative of the
viscosity gradient, so material is accelerated outward in the inner
dead zone boundary (negative viscous gradient) and inward in the
outer dead zone boundary (positive viscous gradient). This modifies
the potential vortensity profile at these transitions, triggering
the RWI (Lyra \& Mac Low 2012). In this study, we investigate the
effect of viscosity on the RWI. For this purpose, we follow the work
of Lovelace and Hohlfeld (2013) by adding the $\alpha$-model
viscosity to their work. Formulation of the problem is given in
section 2. The results and astrophysical implications are given in
section 3, and section 4 is devoted to conclusion and discussion.

\section{Formulation of the Problem}
We use cylindrical coordinate, ($r$,$\theta$,$z$), centered on the
accreting object and make the following standard assumptions:

(1) The equilibrium has the flow velocity $\textbf{u}=u_{\phi}
\hat{e}_\phi=r\Omega(r)\hat{e}_\phi$ where $\Omega(r)$ is angular
velocity at radius $r$. That is, the accretion velocity $u_r$ and
the vertical velocity $u_z$ are assumed negligible compared with
$u_\phi$.

(2) The viscosity is $\nu=\alpha c_s H$ where $c_s$ is the sound
speed, $H$ is the disc height, and $\alpha$ is an important free
parameter between zero (no accretion) and approximately one. In
other words, $\alpha$ is a parameter that is used to model the
unknown sources of angular momentum transport (Pringle 1981).

(3) The gravitational potential $\Phi$ is given by
$\nabla^2\Phi=4\pi G$, where G is the gravitational constant.

(4) The equilibrium flow satisfies $-\Sigma r \Omega^2
=-dP/dr-\Sigma \nabla \Phi$, where $P$ is the vertically integrated
pressure and $\Sigma$ is the surface density.

The perturbed quantities are: density, $\tilde{\Sigma}=\Sigma +
\delta \Sigma(r, \phi, t)$; pressure, $\tilde{P} = P + \delta P(r,
\phi, t)$; flow velocity, $\tilde{\textbf{u}} = \textbf{u} + \delta
\textbf{u}(r, \phi, t)$ where $\delta \textbf{u}=(\delta u_r, \delta
u_{\phi}, 0)$. The equations for the perturbed flow are
\begin{equation}\label{1}
  \frac{D \tilde{\Sigma}}{D t}+\tilde{\Sigma}\nabla\cdot
  \tilde{\textbf{u}}=0,
\end{equation}
\begin{equation}\label{2}
  \frac{D \tilde{\textbf{u}}}{D t}=-\frac{\nabla \tilde{P}}{\tilde{\Sigma}}-\nabla\Phi+\nabla \cdot \Upsilon,
\end{equation}
\begin{equation}\label{5}
 \frac{D S}{D t}=0,
\end{equation}
where $\Upsilon$ is viscous stress tensor, $D/Dt=\partial/\partial
t+\tilde{\textbf{u}}\cdot\nabla$,  and
$S=\tilde{P}/(\tilde{\Sigma})^\gamma$ is the entropy of the disc
matter. Since we consider radially localized modes in the sense that
perturbation extends over a radial region $\Delta r$ around $r_0$
with $(\Delta r)^2\ll r_0^2$ , we use perturbations as follows:
$-f_0\exp(ik_rr+im\phi-i\omega t)$, where amplitude $f_0$ is a
constant, $k_r$ is the radial wavenumber of the perturbation, $m= 0,
1, 2, ...$ is the azimuthal mode number and
$\omega=\omega_r+i\omega_i$ in which for growing modes of interest,
$\omega_i >0$. The perturbations are the first terms of Fourier
expansion and we neglect other terms for simplification.

From equation (1), we have

\begin{equation}\label{4}
 \Delta \omega \delta \Sigma=\left(k_r\Sigma-i\frac{\partial \Sigma}{\partial r}\right)\delta u_r
 +k_\phi \Sigma \delta u_\phi,
\end{equation}
where
\begin{equation}\label{4}
 \Delta \omega(r)=\omega-m\Omega(r).
\end{equation}
From equation (2) we have
\begin{equation}\label{4}
 i\Delta \omega \delta u_r+2\Omega\delta
 u_{\phi}=\frac{ik_r}{\Sigma}\delta P
 -\frac{\delta \Sigma}{\Sigma^2}\frac{d P}{d r}+ik_r\delta \Phi,
\end{equation}
\begin{equation}\label{4}
 i\Delta \omega \delta u_{\phi}-\frac{\Omega_r^2}{2\Omega}\delta
 u_r=ik_{\phi}\frac{\delta P}{\Sigma}+ik_\phi\delta \Phi-\nu
 \left(\frac{ik_r}{r}\delta u_{\phi}-\frac{\delta u_\phi}{r^2}-k_r^2\delta u_{\phi}+\frac{2ik_{\phi}}{r}\delta u_r-k_{\phi}^2\delta
 u_{\phi}\right).
 \end{equation}
Here, $\Omega_r=[r^{-3}d(r^4\Omega^2)/dr]^\frac{1}{2}$ is the radial
epicyclic frequency, and $k_\phi=m/r$ is the azimuthal wavenumber.
For an approximately Keplerian disc, $\Omega_r\approx\Omega$. From
equation (3) and the definition of entropy, we have

\begin{equation}\label{11}
    \delta P=c_s^2\delta \Sigma-\frac{i\Sigma c_s^2}{\Delta \omega
    L_s}\delta u_r,
\end{equation}
where, $c_s = (dP/d\Sigma)^{1/2}_S$ is the effective sound speed and
$L^{-1}_S=\gamma^{-1} d \ln(S)/dr$ with $L_S$ the length-scale of
the entropy variation in the disc. To simplify the subsequent
calculations, we consider the homentropic case where
$L_S\rightarrow\infty$ (Lovelace \& Hohlfeld 2013). For an
approximately Keplerian disc with $\Omega_r\approx \Omega$, we can
neglect of $dP/dr$ in equation (6). In this case, we can rewrite
equations (6) and (7) as follows
\begin{equation}\label{4}
 i\Delta \omega \delta u_r+2\Omega\delta u_{\phi}=ik_r\delta \Psi
 ,
\end{equation}

\begin{equation}\label{4}
  A_0 \delta u_{\phi}-A_1\delta u_r=ik_{\phi}\delta \Psi,
\end{equation}
where
\begin{equation}\label{15}
\nonumber    A_0=\left(i\Delta \omega +\frac{i\nu
k_r}{r}-\frac{\nu}{r^2}-\nu (k_r^2+k_\phi^2)\right), \quad
A_1=\left(\frac{\Omega_r^2}{2\Omega}
 -\frac{2i\nu k_{\phi}}{r}\right),\\
 \nonumber \quad \delta \Psi=c_s^2\frac{\delta \Sigma}{\Sigma}+\delta \Phi.\\
\end{equation}
Equations (9) and (10) can be solved to give
\begin{equation}\label{16}
  \delta  u_r=\frac{\delta \Psi}{2\Omega A_1+iA_0\Delta\omega}\left(iA_0k_r-2ik_{\phi}\Omega
  \right),
\end{equation}
\begin{equation}\label{17}
 \delta   u_{\phi}=\frac{\delta \Psi}{2\Omega A_1+iA_0\Delta\omega}\left(iA_1k_r-k_{\phi}\Delta\omega
 \right).
\end{equation}
By substituting equations (12) and (13) into (4) we obtain
\begin{equation}\label{20}
 (2\Omega A_1+iA_0\Delta\omega)\Delta \omega
\frac{\delta
\Sigma}{\Sigma}=i\left(k_r-\frac{i}{\Sigma}\frac{\partial
\Sigma}{\partial r}\right)\left(A_0k_r-2k_{\phi}\Omega \right)\delta
\Psi+
 i\left(A_1k_\phi
k_r+ik_{\phi}^2\Delta\omega \right)\delta \Psi.
\end{equation}
The perturbation of the gravitational potential is give by
\begin{equation}\label{28}
    \nabla^2\Phi=4\pi G \delta \Sigma \delta(z).
\end{equation}
The WKBJ solution of equation (15) gives $\delta \Phi=-2\pi
G\Sigma/\mid\textbf{k}\mid^2$ where $\textbf{k}=k_r \hat{e}_r+k_\phi
\hat{e}_\phi$ (Lovelace \& Hohlfeld 2013). The WKBJ approximation
allows performing a local perturbation stability analysis of the
perturbed disc (no need to exact boundary conditions because the
analysis is local and not global). The main simplification
introduced by the WKBJ approximation mathematically is that the
Poisson equation for the perturbation becomes a (local) algebraic
relation between the surface density of the spiral wave and its
potential. Therefore, for $\delta \Psi$, we can write
\begin{equation}\label{29}
    \delta \Psi=(1-\frac{2k_c^2}{\mid\textbf{k}\mid^2})c_s^2\frac{\delta
    \Sigma}{\Sigma},
\end{equation}
where
\begin{equation}\label{30}
    k_c=\left(\frac{\pi G \Sigma}{c_s^2}\right)^\frac{1}{2},
\end{equation}
is a characteristic wavenumber. From equation (14) and (16), we can
write dispersion relation as follows
\begin{equation}\label{20}
 (2\Omega A_1+iA_0\Delta\omega)\Delta \omega
=\left(c_s^2-\frac{2k_c^2c_s^2}{\mid\textbf{k}\mid^2}\right)\left[\left(k_r-\frac{i}{\Sigma}\frac{\partial
\Sigma}{\partial r}\right)\left(iA_0k_r-2ik_{\phi}\Omega \right)+
 \left(iA_1k_\phi
k_r-k_{\phi}^2\Delta\omega \right)\right].
\end{equation}

For obtaining $\partial \Sigma/\partial r$, we consider a density
bump in a thin accretion disc of the form (see Lovelace et al. 1999)
\begin{equation}\label{21}
    \frac{\Sigma(r)}{\Sigma_0}=1+(F-1)\exp\left[-\frac{(r-r_0)^2}{2\Delta^2}\right],
\end{equation}
where the subscript '0' implies that the quantities evaluated at $r
= r_0$ and $\Sigma_0$, $F$ and $\Delta$ are constants, respectively.
Following Yu \& Lai (2013), we chose $0.4 < r/r_0 < 1.6$, $F= 1.2$
and $\Delta = 0.05r_0$ through this paper. We further assume $c_s =
0.1r_0\Omega_ 0 =$ constant.

For axisymmetric perturbations $(k_\phi = 0 = m)$ of smooth disc
without viscosity $(\nu=0)$ and with neglecting the radial variation
of $\Sigma$, we can write (Safranov 1960; Toomre 1964)
\begin{equation}\label{31}
    \omega^2=\Omega_r^2+k_r^2c_s^2-2k_c^2c_s^2.
\end{equation}
The minimum of $[\omega(k_r)]^2$ occurs at $k_r = k_c$ where
$\omega^2=\Omega^2-k_c^2c_s^2$. Therefore, if we define
$k_{c*}=\Omega_r/c_s$, then for $k_c < k_{c*}$, the minimum of
$\omega^2$ is positive and the axisymmetric perturbations are
stable. Conversely for $k_c>k_{c*}$, the perturbations are unstable.
With (Lovelace \& Hohlfeld 2013)
\begin{equation}\label{32}
    Q=\frac{k_{c*}}{k_c}=\frac{\Omega_rc_s}{\pi G \Sigma},
\end{equation}
the axisymmetric perturbations are stable (unstable) for $Q > 1
\quad (Q < 1)$ (Toomre 1964). The minimum value of the squared
frequency is $\omega(kr)^2 = (k_{c*}c_s)^2(1- Q^{-2})$.

To obtain non-axisymmetric modes, by choosing following
dimensionless parameters and variables
\begin{eqnarray}\label{21}
\nonumber  &&R=\frac{r}{r_0}, \quad
\varpi=\frac{r_0\Delta\omega}{c_s},\quad\gamma=\frac{r_0\Omega}{c_s},\quad\gamma_r=\frac{r_0\Omega_r}{c_s},
\quad\varrho=\frac{\Sigma}{\Sigma_0},
\quad\mu=\frac{\nu}{r_0c_s},\\
\nonumber &&\quad \sigma_r=k_rr_0,
    \quad \sigma_\phi=k_\phi r_0,\quad\sigma_c=k_cr_0,
   \\
\end{eqnarray}
we can rewrite equation (18) as follows
\begin{equation}\label{25}
    \varpi^3+C_2\varpi^2+C_1\varpi+C_0=0,
\end{equation}
where $C_2$, $C_1$ and $C_0$ are complex coefficients in the
equations below
\begin{equation}\label{22}
C_2=\left(\frac{\mu \sigma_r}{R}+\frac{i\mu}{R^2}+i\mu
(\sigma_r^2+\sigma_\phi^2)\right),
\end{equation}

\begin{equation}\label{20}
C_1
=-\left(1-\frac{2\sigma_c^2}{\sigma_r^2+\sigma_\phi^2}\right)\left[\sigma_{\phi}^2+\left(\sigma_r-\frac{i}{\varrho}\frac{\partial
\varrho}{\partial R}\right)\sigma_r\right]- \left(\gamma_r^2
 -\frac{4i\gamma \mu \sigma_{\phi}}{R}\right),
\end{equation}

\begin{eqnarray}\label{100}
\nonumber
C_0&=&\left(1-\frac{2\sigma_c^2}{\sigma_r^2+\sigma_\phi^2}\right)\\
\nonumber&&\left[\left(\frac{i}{\varrho}\frac{\partial
\varrho}{\partial r}-\sigma_r\right)\left(\left(\frac{\mu
\sigma_r}{R}+\frac{i\mu}{R^2}+i\mu
(\sigma_r^2+\sigma_\phi^2)\right)\sigma_r+2i\sigma_{\phi}\gamma
\right)
 + i\left(\frac{\gamma_r^2}{2\gamma}
 -\frac{2i\mu \sigma_{\phi}}{R}\right)\sigma_\phi \sigma_r\right].\\
\end{eqnarray}
The characteristic equation (23) allows determination of $\varpi$,
and must be solved numerically to determine stable and unstable
modes. For this purpose, we use the Laguerre method (Press et al.
1992) to obtain its three roots.

\section{Results and Astrophysical Implications}
In this section, we consider the results and their astrophysical
implications with an emphasis on the formation of planets through
accretion discs. We investigate the problem with three ranges of
viscosities as follows: high viscosity in the ranges $0.1\leq
\alpha\leq 0.4$, moderate viscosity in the ranges $0.01 \leq\alpha
<0.1$, and low viscosity in the ranges $\alpha<0.01$. In each range,
we choose an arbitrary value of $0.2, 0.08$ and $0.006$ to represent
the high, moderate and low viscosity, respectively. The stable and
unstable regions versus to the azimuthal mode number, $m$, and the
non-dimensional radial wavenumber, $\sigma_r$, are shown in Fig.(1)
in which the unstable regions are depicted by dark lines. The
Fig.(1) shows the effects of viscosity on the stable and unstable
modes so that the perturbations corresponding to small azimuthal
mode numbers may be damped for some non-dimensional radial
wavenumbers. Also, Fig.(2) shows the growth rate of unstable modes
versus to the non-dimensional radial wavenumber.

Fig.(1)-(a) and Fig.(2)-(a) are assigned to low viscosities which
occur in weakly ionized protoplanetary discs around T-Tauri stars.
In this case, the viscosity can be removed from the RWI problem
because it does not have important impact. In other words, the low
viscosity is not able to damp the perturbations of the Rossby waves.

Fig.(1)-(b) and Fig.(2)-(b) are assigned to moderate viscosities.
The results of these figures can be utilized in the study of the RWI
in the partially ionized protoplanetary discs around the stars of a
young binary system (e.g., Hartmann 2007). According to these
figures, we can deduce that the RWI, for some radial wavenumbers and
azimuthal mode numbers ($m<4$), may play an important role in the
planetesimal formation and angular momentum transport through these
protoplanetary discs.

Fig.(1)-(c) and Fig.(2)-(c) are assigned to high viscosities.
Fig.(1)-(c) shows that the stable regions decrease with increasing
the azimuthal mode number, $m$, and totally disappear for $m>7$.
These results can be utilized in the study of the RWI in
fully-ionized protoplanetary discs, dwarf-nova accretion discs and
hot discs around supermassive black holes (e.g., King et al. 2007).
The temperature of hot disc increases the viscosity (e.g., Gholipour
\& Nejad-Asghar 2013) which is able to damp the perturbation of the
Rossby waves (corresponding to small azimuthal mode number for
$m<7$). Also, these figures show that the RWI doesn't have important
role in planetesimal formation and angular momentum transport in
fully-ionized protoplanetary discs, dwarf-nova accretion discs and
hot discs around supermassive black holes.

The growth rate versus to the viscosity coefficient is shown in
Fig.(3). It implies that we have larger growth rate for smaller
viscosity. The growth rate of each plot is found to be not only
dependent on the viscosity coefficient but also dependent on the
azimuthal mode numbers. The growth rate and the area under the curve
increase with decreasing the viscosity coefficient and with
increasing of $m$. Obviously, Fig.(3) shows that for $m=1$ and
$m=2$, the instability doesn't occur for high viscosity unless in
small radial wavenumbers, and the distance between the lines becomes
smaller with decreasing wavenumber. This figure has important
results in obtaining timescale of planetesimal formation by the RWI
in the protoplanetary discs. The rate of planetesimal formation is
faster for a protoplanetary disc with low viscosity than one with
high viscosity. This subject may be considered as a responsible
process in problem of disc dispersal (Bodenhimer 2011).

The non-dimensional critical wavelength
($\tilde{\lambda}_{crit}=\lambda_{crit}/r_0$) versus the viscosity
coefficient is shown in Fig.(4). It explains what would be the
viscosity necessary to prevent the RWI from growing in a thin
accretion disc. Since the long and intermediate wavelengths
($\lambda>r_0$) are not important in planetesimal formation process,
the maximum of non-dimensional critical wavelength is considered
equal to one. For example, if we assume $\alpha>0.1$ for a hot disc
(such as supermassive black hole accretion disc), the planetesimal
formation by the RWI has less chance for occurrence in $m=1$ than
$m=2$ or $m=3$. Thus, we can neglect the contribution of $m=1$ for a
hot disc around supermassive black hole with $\alpha>0.22$ (see
Fig.(4)). In this case, the larger numbers of the azimuthal mode
number may be contributed.

\section{Conclusion and Discussion}
In this paper, we have carried out linear analysis of the RWI in
accretion discs including the viscosity. The RWI may play an
important role in planetesimal formation and angular momentum
transport in weakly ionized protoplanetary discs. In the study of
Lovelace and Hohlfeld (2013), the viscosity was ignored and the
self-gravity was emphasized, but we added the viscosity in their
self gravitating work. The viscosity complicates the problem, thus
we used the numerical methods to investigate the stable and unstable
modes. The results show that the occurrence of the RWI is related to
the value of viscosity. Since the hot discs have higher viscosity
than other discs (e.g., Gholipour \& Nejad-Asghar 2013; King et al.
2007), the high viscosity is able to damp the perturbations
corresponding to small azimuthal mode numbers. Our results indicate
that the RWI may play more important role in weakly ionized thin
disc than the fully or partially ionized one. In other words, the
contribution of low viscosity may be ignored in the RWI. Here, we
also considered the results of King et al. (2007) which considered
observational and theoretical estimates of the accretion disc
viscosity coefficient. They found that in thin, fully ionized discs,
the best observational evidence suggests a typical range $\alpha
\sim 0.1-0.4$, whereas the relevant numerical simulations tend to
derive estimates for $\alpha$ which are an order of magnitude
smaller. To compare with simulation works, we can mention the work
of Varniere and Tagger (2006). They found the consequences on the
disc dynamics of the presence of dead zone, where the transport of
matter and angular momentum (with a turbulent viscosity) is
significantly lower than elsewhere in the disc. Our results are in
agreement with this simulation work.

The results show the RWI can be damped in the hot discs so that it
doesn't have an important role in the planetesimal formation in
these discs. In this case, other processes can be considered
according to the high viscosity. For example, Gholipour and
Nejad-Asghar (2013) introduced viscothermal instability that the
viscosity of hot disc causes thermal instability which may lead to
planetesimal formation through discs. Also, the results of this
paper can be considered in problem of disc dispersal. Numerous
mechanisms, most still under investigation, have been proposed to
explain the fact that observational evidence for the presence of
discs around newly formed stars disappears once the stars reach an
age of 1-10 Myr (Bodenhimer 2011). Our results predict that the rate
of planetesimal formation is faster for a protoplanetary disc with
low viscosity than a protoplanetary disc with high viscosity (see
Fig.(3)). Since the discs around newly formed stars have low
viscosity, $\alpha<0.01$ (Bodenhimer 2011), we expect the disc
disperses faster around protostar than around supermassive black
holes. Briefly, the results show that the important quantities
determining the stability/instability are: (1) the viscosity at
region of the density bump (2) the azimuthal mode number (3) the
radial wavelength. These results may explain why the formation of
planets is impossible in some regions.

\section*{Acknowledgments}
We thank the referee and Prof. Richard V. E. Lovelace for very
constructive and valuable comments that helped us to improve the
initial version of the paper.



\clearpage
\begin{figure}
\epsscale{.5} \center \plotone{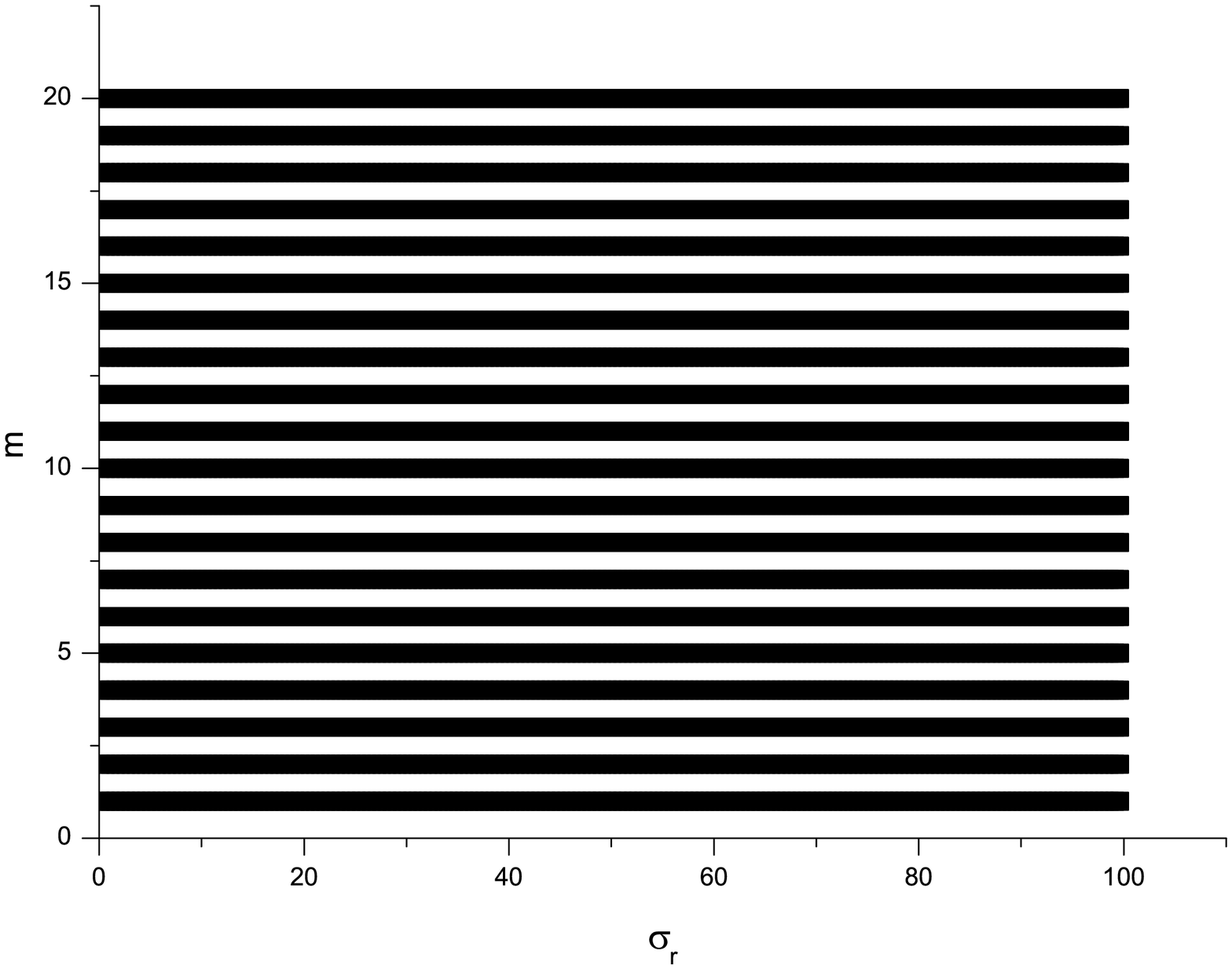}\\{(a)}\\
\epsscale{.5} \center \plotone{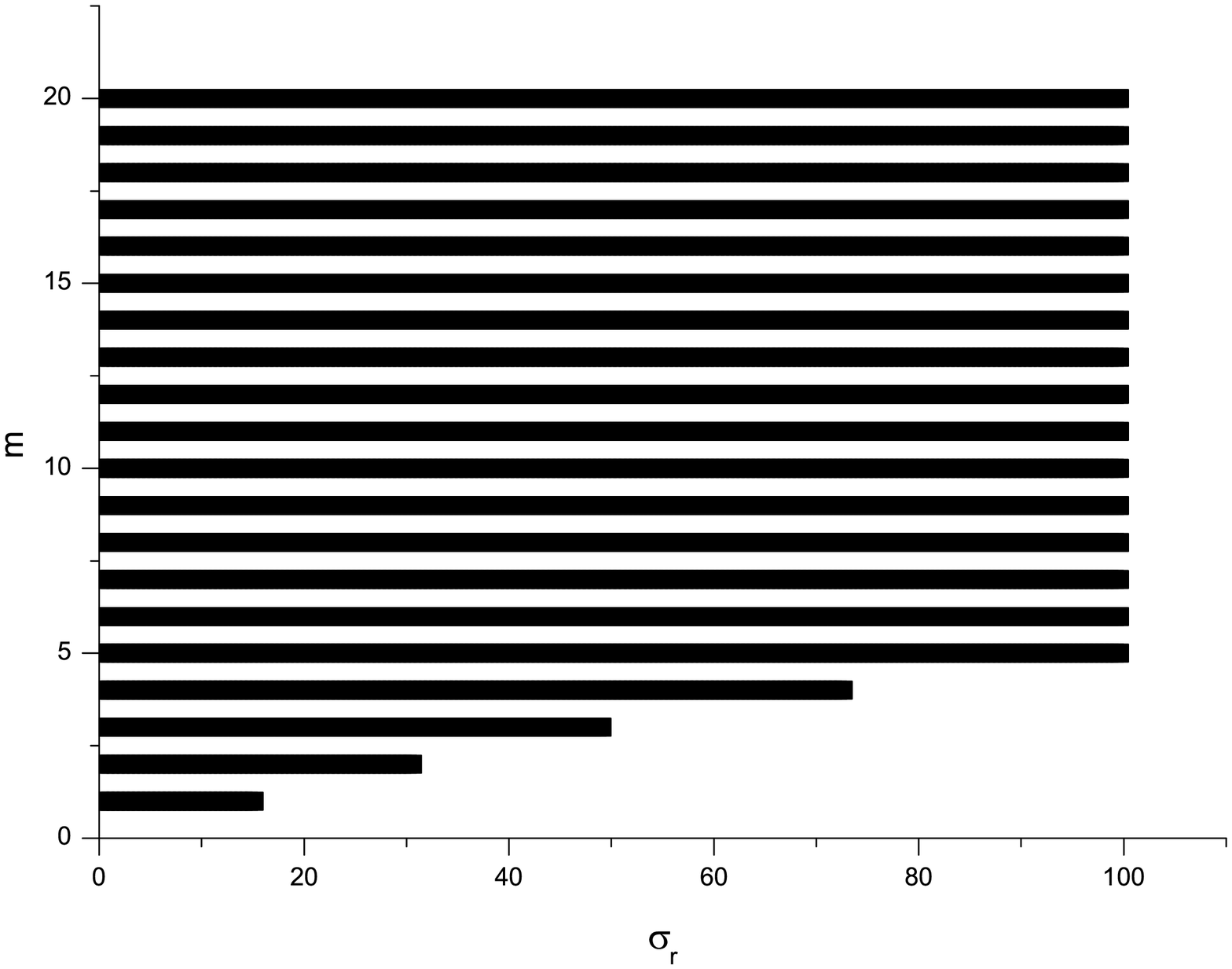}\\{(b)}\\
\epsscale{.5} \center \plotone{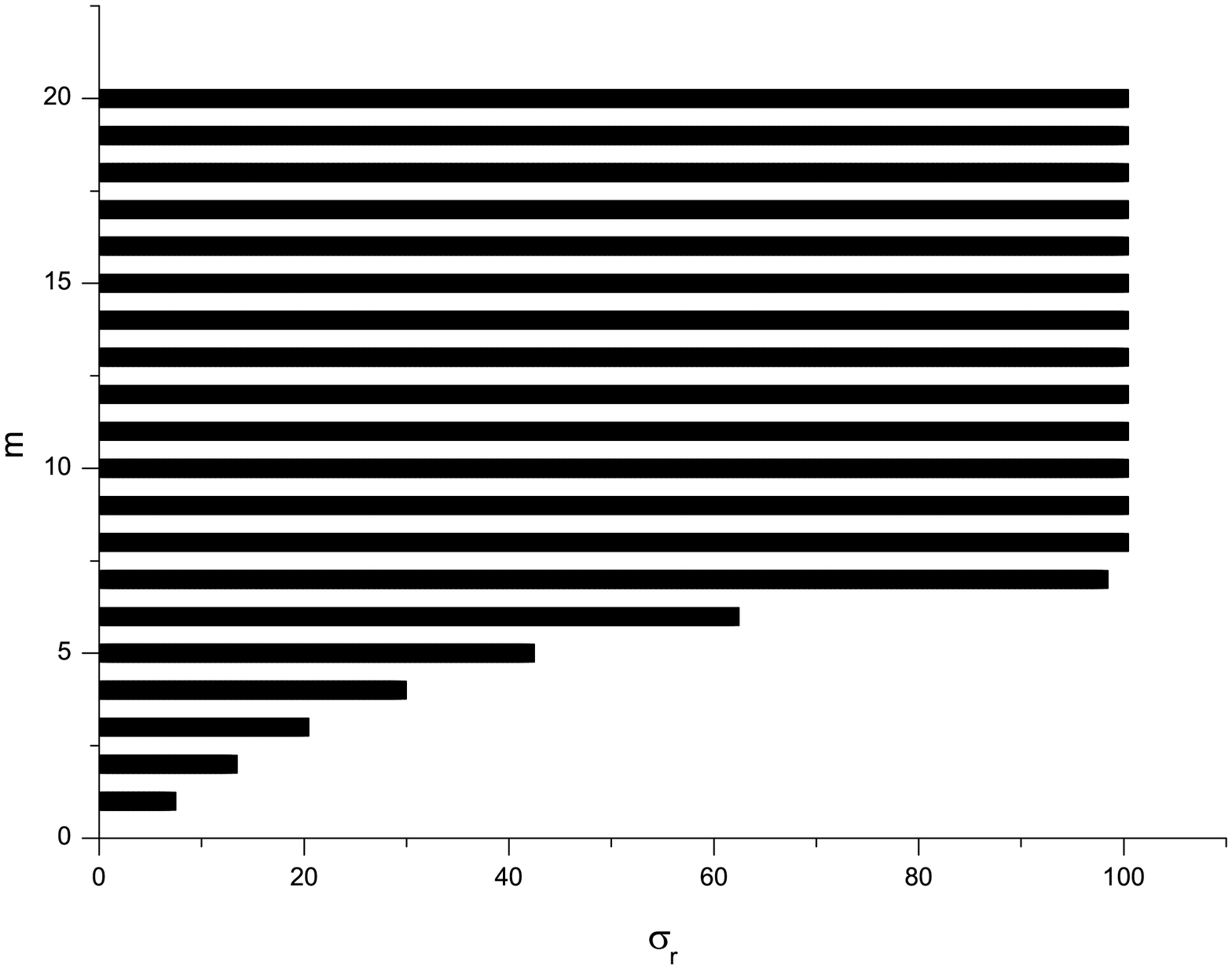}\\{(c)}\\ \caption{The
stable and unstable regions versus to the azimuthal mode number,
$m$, and the non-dimensional radial wavenumber, $\sigma_r$, are
shown in Fig.(1) in which the unstable regions are depicted by dark
lines for cases with (a) $\alpha=0.006$, (b) $\alpha=0.08$ and (c)
$\alpha=0.2$.}
\end{figure}
\clearpage

\begin{figure}
\epsscale{.6} \center \plotone{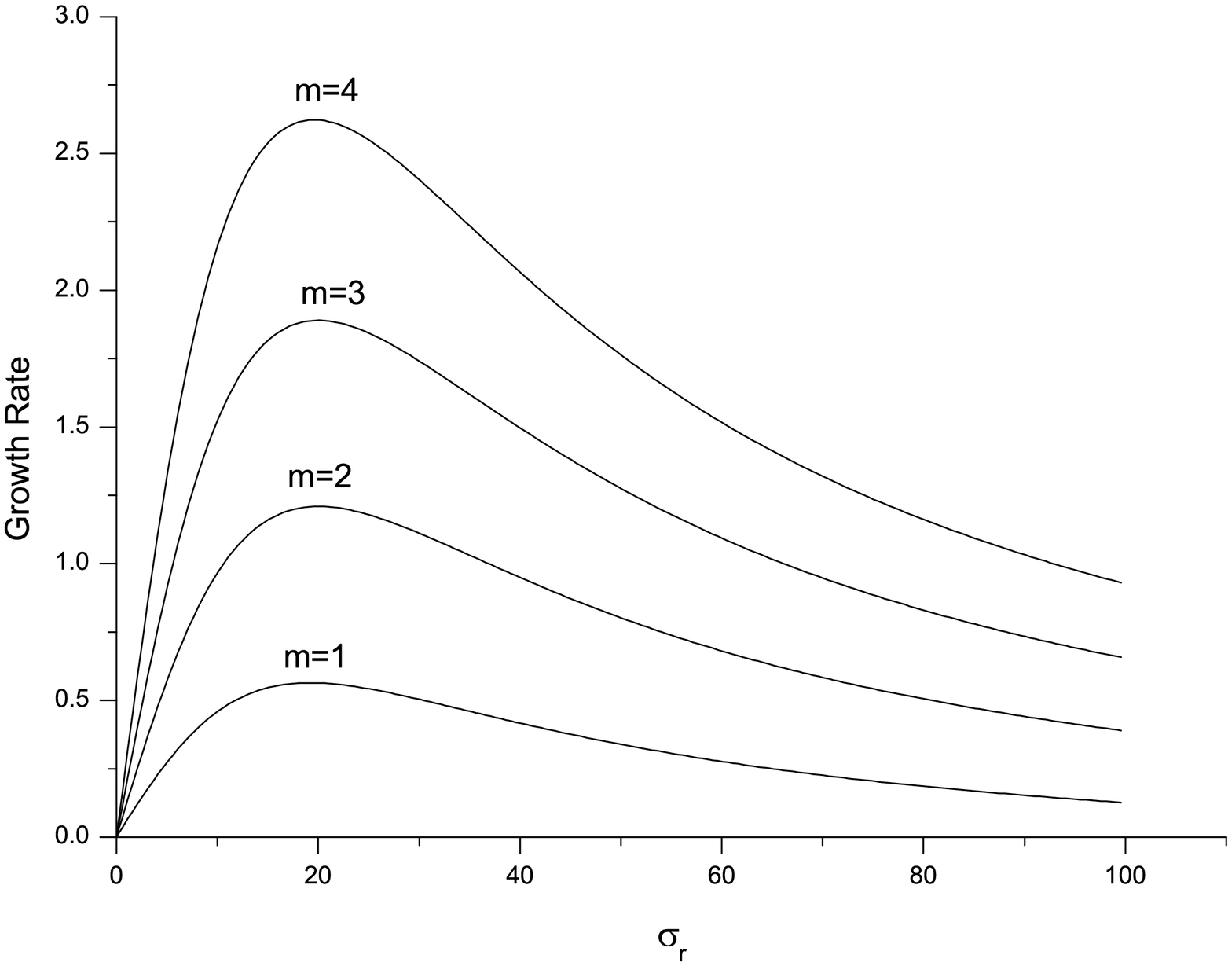}\\{(a)}\\
\epsscale{.6} \center \plotone{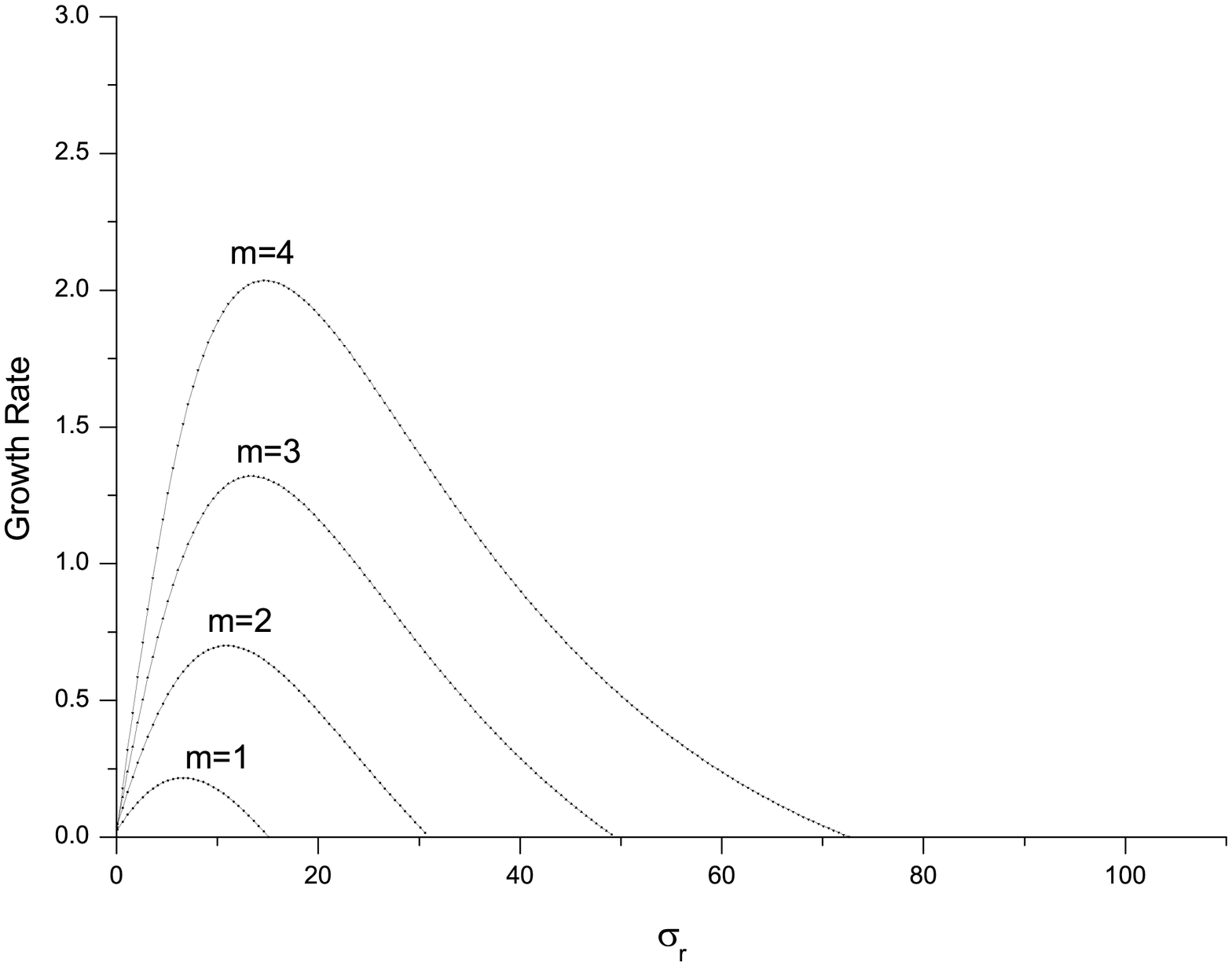}\\{(b)}\\
\epsscale{.6} \center \plotone{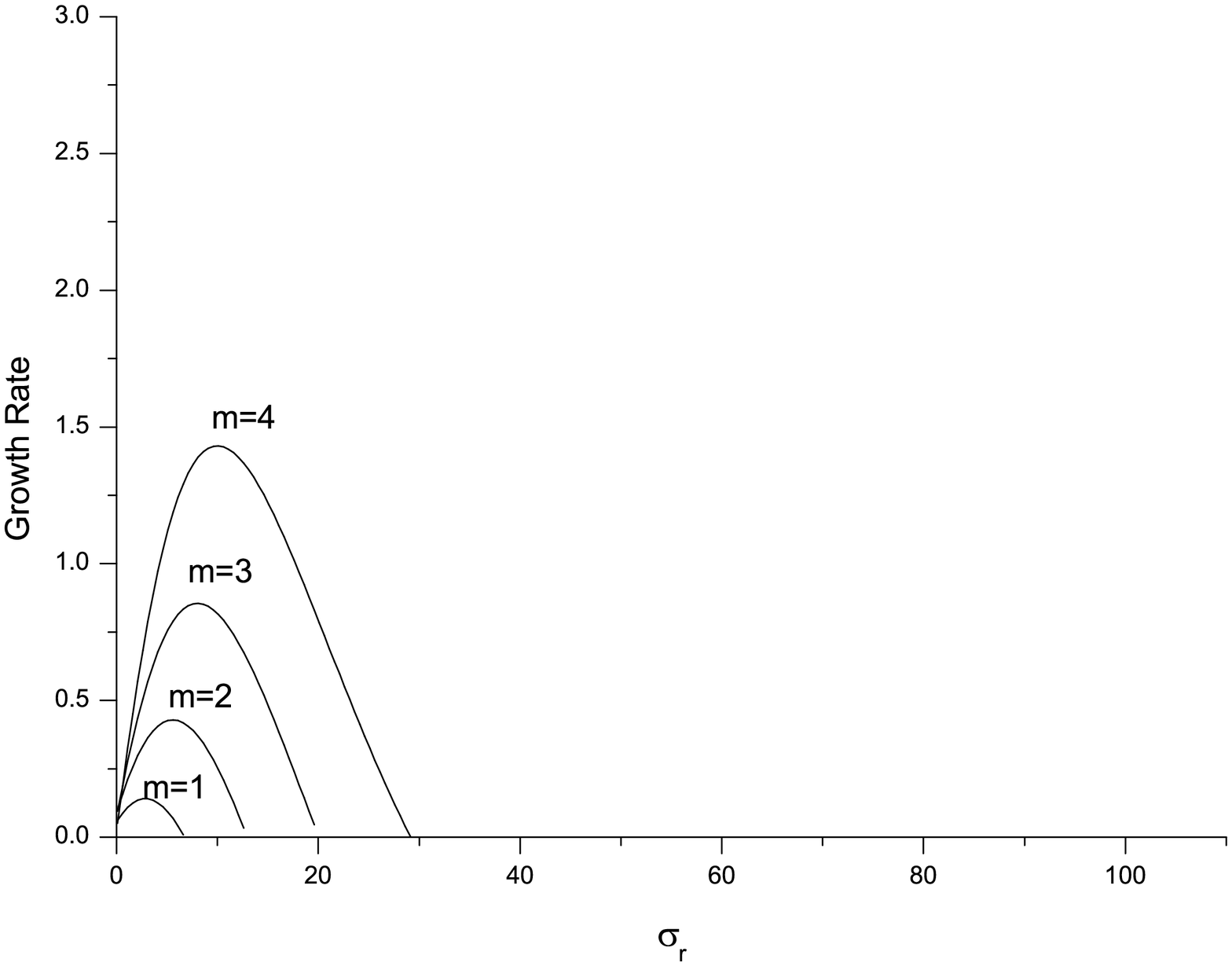}\\{(c)}\\
 \caption{The growth rate versus
to the non-dimensional radial wavenumber ($\sigma_r$) for cases with
$m=4, 3, 2, 1$ and (a) $\alpha=0.006$, (b) $\alpha=0.08$ and (c)
$\alpha=0.2$.}
\end{figure}
\clearpage

\begin{figure}
\epsscale{.56} \center \plotone{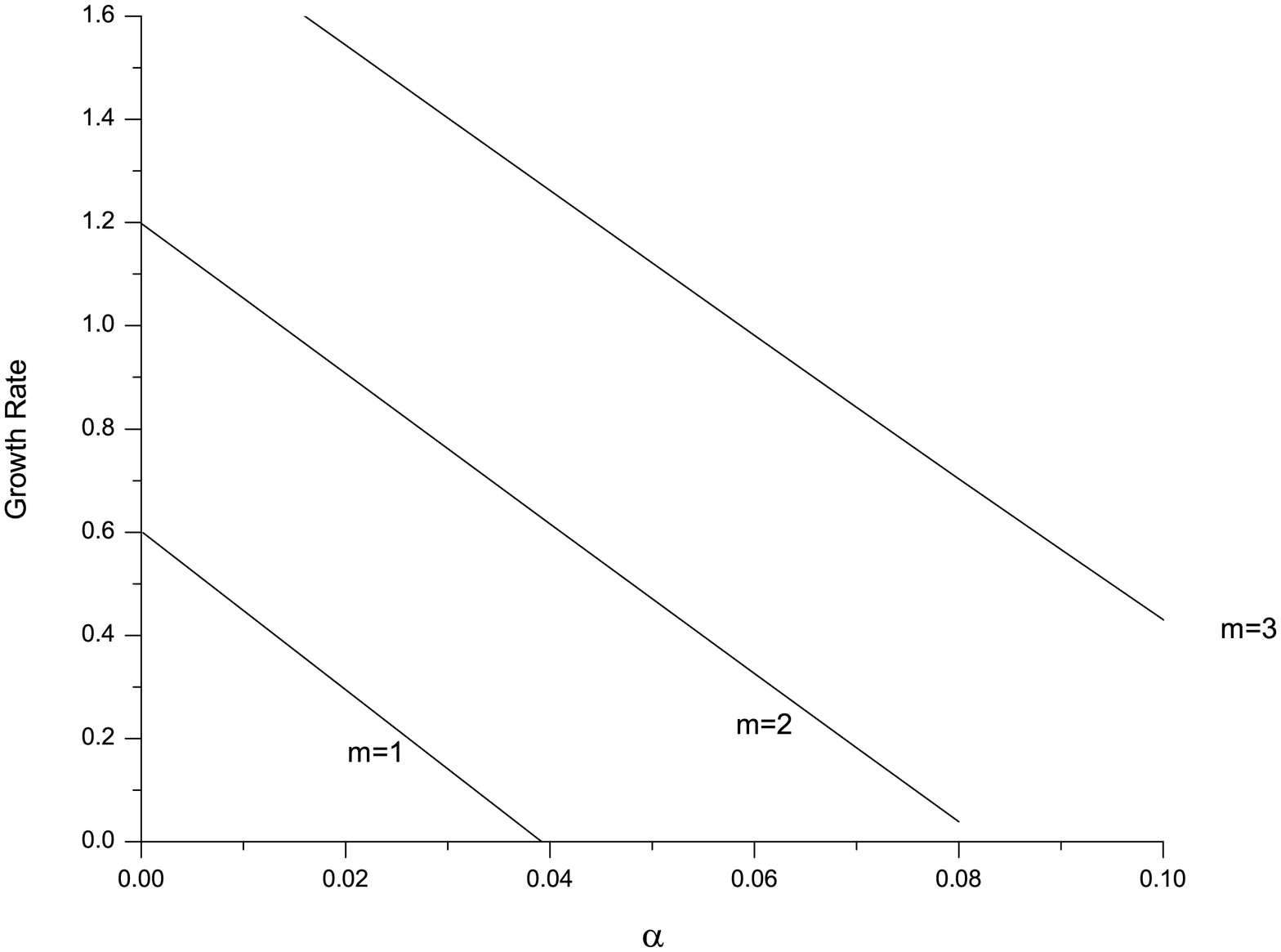}\\{(a)}\\
\epsscale{.56} \center \plotone{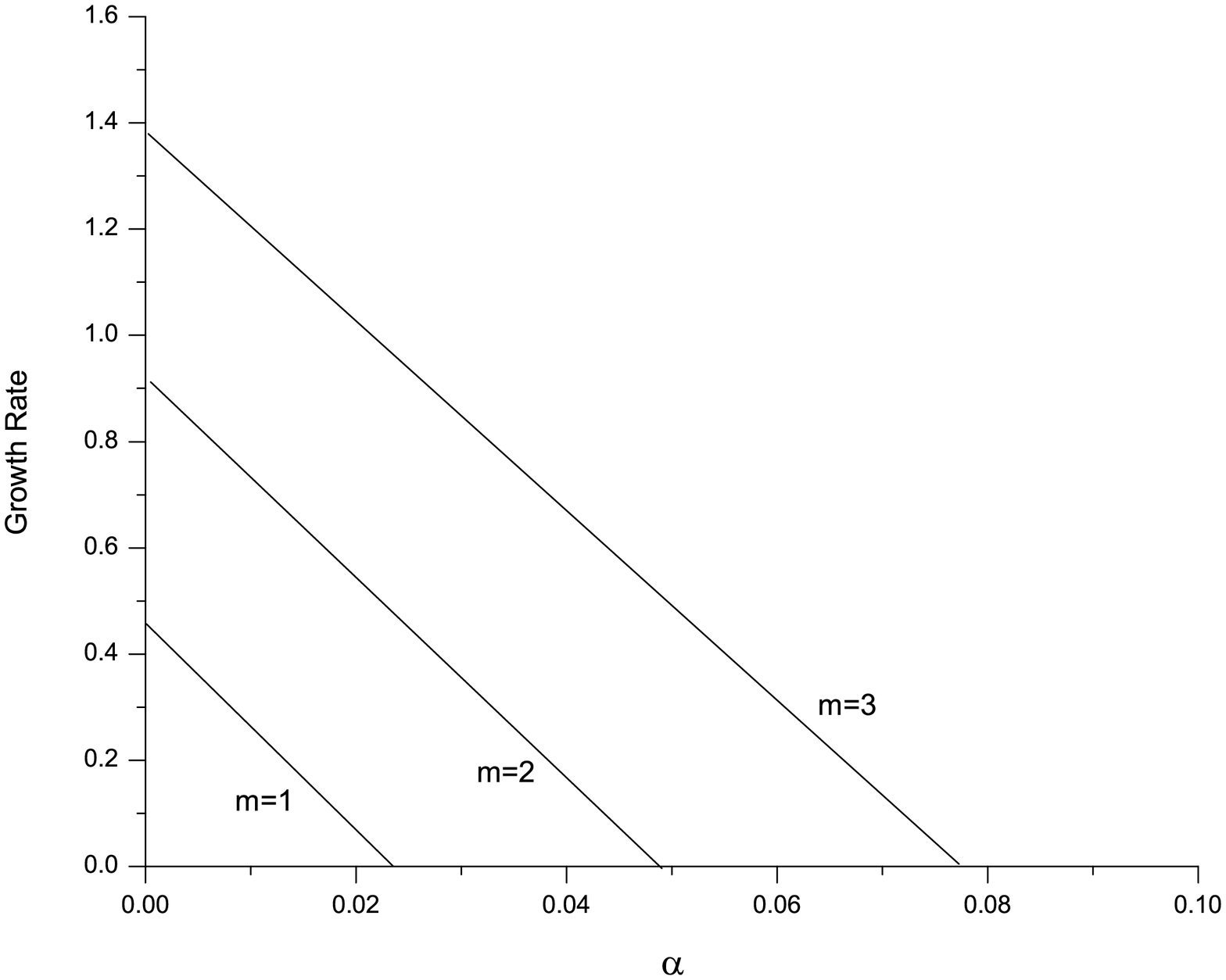}\\{(b)}\\
\epsscale{.56} \center \plotone{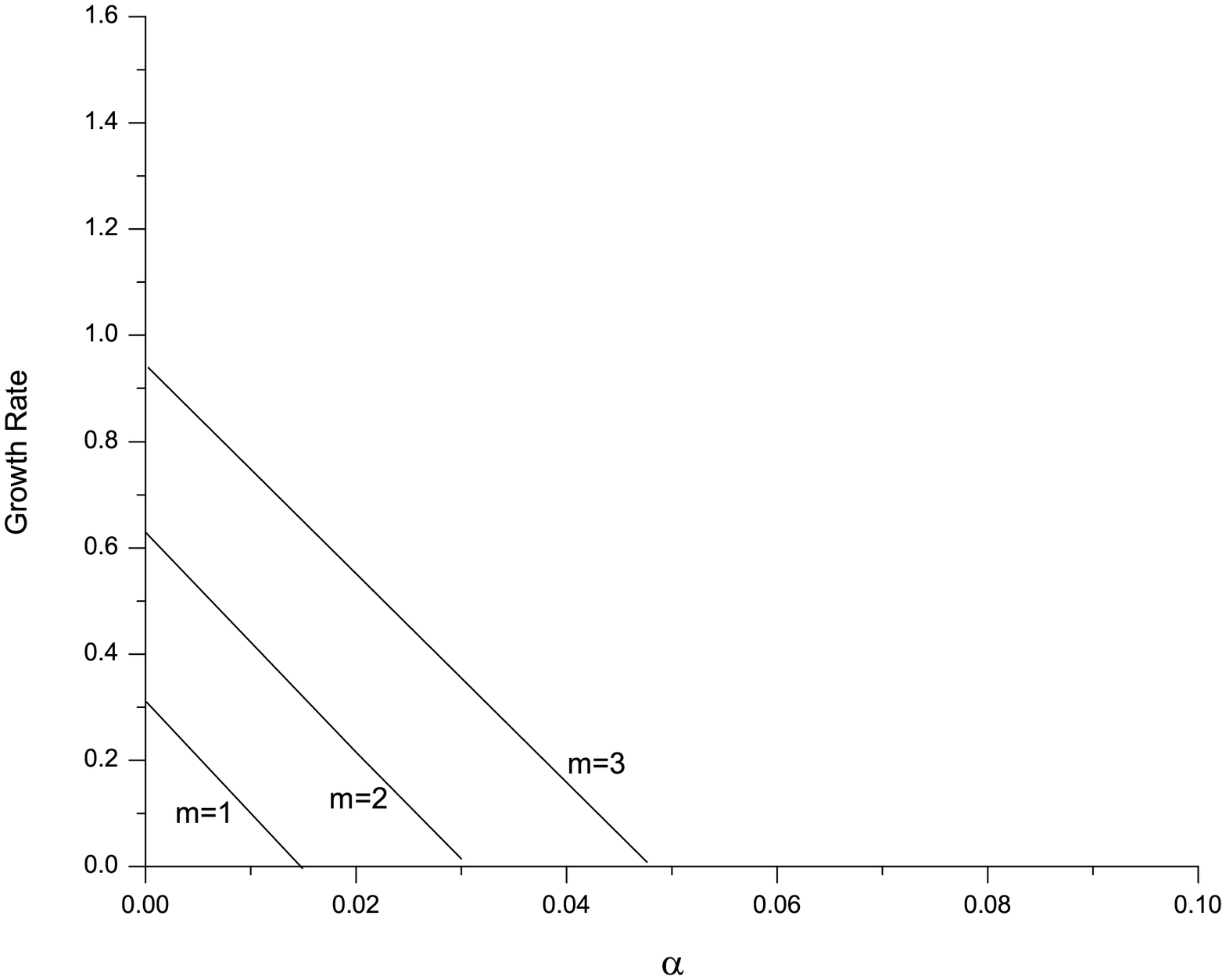}\\{(c)}\\
 \caption{The growth rate versus
the viscosity coefficient for cases with $m=3, 2, 1$ and (a)
$\sigma_r=30$, (b) $\sigma_r=50$, (c) $\sigma_r=80$.}
\end{figure}
\clearpage
\begin{figure} \epsscale{0.9} \plotone{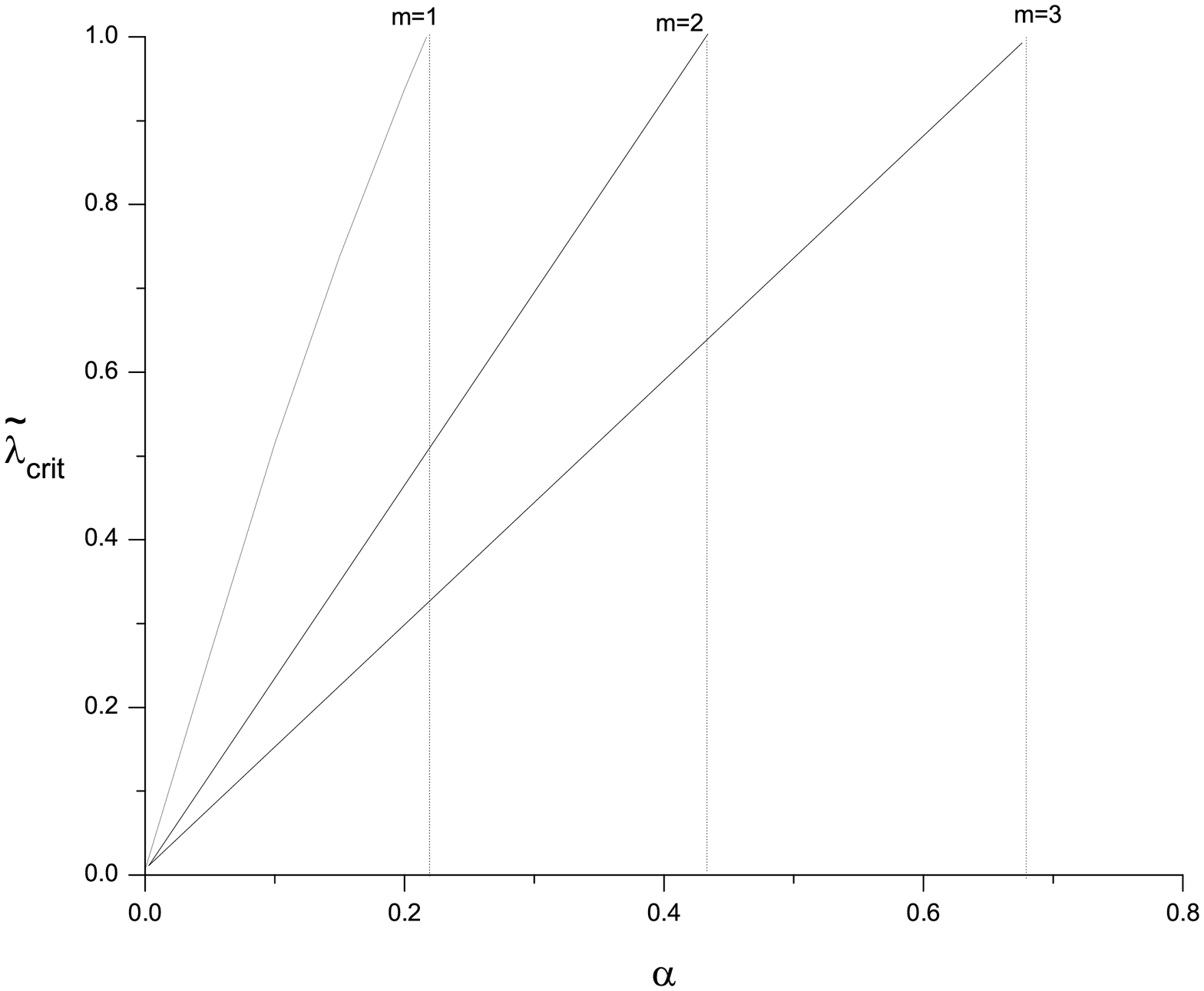}
\caption{The non-dimensional critical wavelength
($\tilde{\lambda}_{crit}=\lambda_{crit}/r_0$) versus to $\alpha$
(the viscosity coefficient) with $m=3, 2, 1$.}
\end{figure}
\end{document}